\pdfoutput=1
 \documentclass[aps, prapplied,twocolumn,showpacs,superscriptaddress,preprintnumbers, longbibliography,nofootinbib,floatfix]{revtex4-2}
\usepackage[normalem]{ulem}
\usepackage{amsmath}
\usepackage{amssymb}
\usepackage{physics}  % http://mirrors.ibiblio.org/CTAN/macros/latex/contrib/physics/physics.pdf
\usepackage{epsfig}

\usepackage{url} 

\usepackage{graphicx}
\usepackage{hyperref}
\usepackage{cleveref}
\crefname{figure}{Fig.}{Figs.}
\Crefname{figure}{Fig.}{Figs.}
\crefname{equation}{Eq.}{Eqs.}
\crefname{table}{Table}{Tables}
\Crefname{equation}{Eq.}{Eqs.}
\crefname{section}{Sec.}{Secs.}
\Crefname{section}{Sec.}{Secs.}
\usepackage{tabu}
\usepackage{boldline}
\usepackage{xspace}
\usepackage{slashed}
\usepackage{multirow}
\usepackage{booktabs}
\usepackage{diagbox}
\usepackage{tabularx}
\usepackage{comment}
\usepackage{floatrow}
\newfloatcommand{capbtabbox}{table}[][\FBwidth]
\usepackage{color}
\usepackage[normal]{subfigure}
\usepackage{enumitem}
\usepackage[utf8]{inputenc}
\usepackage[english]{babel}
\usepackage{colortbl}
\usepackage{upgreek}
\usepackage[separate-uncertainty=true]{siunitx}
\usepackage[dvipsnames]{xcolor}
\AtBeginDocument{\RenewCommandCopy\qty\SI}

\newcommand{\ntraps}{1787}
\newcommand\sjt{\bgroup\markoverwith{\textcolor{blue}{\rule[0.5ex]{2pt}{0.4pt}}}\ULon}

\DeclareCaptionJustification{justified}{\justified}

\renewcommand{\arraystretch}{1.15}
\DeclareSIUnit{\electron}{e\textsuperscript{-}}
\DeclareSIUnit{\bar}{bar}
\DeclareSIUnit{\pix}{\textup{pixel}}
\DeclareSIUnit{\day}{\textup{day}}
\DeclareSIUnit{\epd}{\electron\per\pix\per\day}
\DeclareSIUnit{\ADU}{\text{A.D.U.}}

\usepackage{lipsum} 

\begin{document}

\preprint{FERMILAB-PUB-26-0763-PPD}

\title{Long-Lived Charge Trap Characterization for Rare-Event Searches with Skipper-CCDs}

% ------------------------------------------------------
% ------------------------------------------------------

\author{Ansh Desai}
\affiliation{\normalsize\it 
Institute for Fundamental Science, University of Oregon, Eugene, OR 97403, USA}

\author{Ana M. Botti}
\affiliation{\normalsize\it 
Fermi National Accelerator Laboratory, PO Box 500, Batavia IL, 60510, USA}
\affiliation{\normalsize\it Kavli Institute for Cosmological Physics, University of Chicago, Chicago, IL 60637, USA}

\author{Brenda A. Cervantes-Vergara}
\affiliation{\normalsize\it 
University of Chicago, Chicago, IL 60637, USA}

\author{Rouven Essig}
\affiliation{\normalsize\it 
C.N.~Yang Institute for Theoretical Physics, Stony Brook University, Stony Brook, NY 11794, USA}

\author{Stephen E. Holland}
\affiliation{\normalsize\it Lawrence Berkeley National Laboratory, One Cyclotron Road, Berkeley, California 94720, USA}

\author{Ian Lawson}
\affiliation{\normalsize\it SNOLAB, Lively, ON P3Y 1N2, Canada}
\affiliation{\normalsize\it School of Natural Sciences, Laurentian University, Sudbury, P3E 2C6, Canada} 

\author{Steffon Luoma}
\affiliation{\normalsize\it SNOLAB, Lively, ON P3Y 1N2, Canada}

\author{Santiago Perez}
\affiliation{\normalsize\it University of Chicago, Chicago, IL 60637, USA}

\author{Javier Tiffenberg}
\affiliation{\normalsize\it 
Fermi National Accelerator Laboratory, PO Box 500, Batavia IL, 60510, USA}
\affiliation{\normalsize\it 
Universidad de Buenos Aires, Facultad de Ciencias Exactas y Naturales, Departamento de Física, Buenos Aires, Argentina}

\author{Sho Uemura}
\affiliation{\normalsize\it 
Fermi National Accelerator Laboratory, PO Box 500, Batavia IL, 60510, USA}

\author{Yikai Wu}
\affiliation{\normalsize\it 
C.N.~Yang Institute for Theoretical Physics, Stony Brook University, Stony Brook, NY 11794, USA}
\affiliation{\normalsize\it 
Department of Physics and Astronomy, Stony Brook University, Stony Brook, NY 11794, USA}

\author{Tien-Tien Yu}
\affiliation{\normalsize\it 
Institute for Fundamental Science, University of Oregon, Eugene, OR 97403, USA}
% ------------------------------------------------------
% ------------------------------------------------------

\begin{abstract}
\noindent Skipper Charge-Coupled Devices (Skipper-CCDs) are detectors with sub-electron resolution, allowing them to play a critical role in rare-event searches. 
A major challenge in experiments using these detectors is understanding and mitigating low-energy backgrounds which impact the single-electron rate. 
One potential component of the single-electron rate could arise from single-electron charge traps caused by lattice defects. 
In this work we use a new in-situ technique for generating uniform charge known as ``pixel-pumping'' across the CCD to study single-electron traps in a SENSEI sensor placed underground in the MINOS cavern at Fermilab, which enables sensitivity to long-lived traps with emission times $>10^3$ seconds, a regime that was previously not accessible. 
We fully characterize \ntraps\;traps and use this information to estimate their contribution to the single-electron background in SENSEI.
Through simulation, we find the characterized population of traps does not significantly impact the single-electron rate.
We further identify the energies and cross-sections for traps to which this analysis is sensitive, and compare this with the values that could impact the single-electron rate. This analysis probes a large fraction of this region.
However, we also see that a sufficiently numerous hidden population could contribute significantly to the single-electron rate, and new measurement techniques will have to be devised in order to fully rule out this population.

\end{abstract}

\maketitle

\section{Traps and single-electron backgrounds}\label{sec:intro}
Charge-Coupled Devices (CCDs) are pixelated silicon detectors~\cite{Smith2010, Holland:2003, Janesick1990,Wen1974} widely used in astronomy and rare-event searches. 
Skipper-CCDs are enhanced versions of CCDs that can perform repeated non-destructive measurements of output charge and consequently achieve sub-electron resolution~\cite{Tiffenberg:2017aac}, enabling sensitivity to particles that produce one or a few electron-hole pairs when they interact with the bulk silicon.
This has allowed the SENSEI and DAMIC experiments to achieve world-leading results in the search for light dark matter~\cite{Abramoff:2019dfb, Sensei2020, SENSEI:2023zdf,DAMIC-M:2023gxo,DAMIC-M:2025luv}. 
These experiments are currently limited by low-energy backgrounds, particularly contributions to the single-electron rate (SER). 
These backgrounds may arise from environmental effects~\cite{Du_2022,Du:2023soy,Sensei2020,SENSEI:2023zdf,SENSEI:2024yyt} or intrinsic processes within the sensor~\cite{senseiSEE,ampLight}, and characterizing them is crucial for maximizing the sensitivity of these and other experiments. 
One possible source of the SER is charge deferred by single-electron trap emission, which we investigate in detail in this work.

It is well known that defects can occur in the silicon lattice within the buried channel of a CCD introducing intermediate energy states that could trap and release charge after a characteristic emission time determined by the properties of the defect or contaminant~\cite{claeys_metal_2018,pichler_intrinsic_2004,chroneos_oxygen_2015,wang_phosphorousvacancyoxygen_2013,collins_properties_1957,pintiliea_radiation_2009}. 
If traps have sufficiently long emission times, they can produce uncorrelated single-electron events, therefore contributing to the SER. 
Results from the OSCURA collaboration~\cite{oscurasensors, trapsOscura} indicated that contaminants in the silicon lattice of Skipper-CCDs could contribute to the measured SER, motivating dedicated studies in SENSEI sensors. 
More recent measurements performed at LAMBDA Laboratory at the University of Buenos Aires on a CCD fabricated from the same wafer as science-grade SENSEI CCDs~\cite{Brusco:2025kus} suggested that current analysis strategies may already account for charge deferred by these traps. 
However, because these measurements were taken at the surface, high rates of environmental and cosmic backgrounds limited the sensitivity to a restricted range of trap lifetimes. 
In order to probe very long-lived traps (traps that emit charge carriers several days or even months after the initial trapping) at the usual operating temperature of $\sim135$~K, it is necessary to take longer pumping measurements at higher operating temperatures, which must be done in a shielded underground environment. 
In this work we describe such a measurement, which we perform with a SENSEI CCD $\sim100$~m underground in the MINOS cavern at Fermi National Accelerator Laboratory (FNAL).

The Skipper-CCDs used in this study are made of high-resistivity silicon with an active area of $6144\times1024$ pixels, each of size 15 $\times$ 15 $\mu$m. 
Each pixel is covered by three overlapping gates (called ``clocks'' or ``phases'') which are held at different voltages to create a potential well to hold charge. 
Clocking these voltages between high and low states in sequence transfers charge from pixel to pixel. 
Our CCDs were designed by Lawrence Berkeley National Laboratory and fabricated at Teledyne/DALSA. They are p-channel devices and therefore charge carriers are holes and not electrons. 
We use ``electron'' to refer to these holes below. 
During readout, charge in pixels of the active region is shifted row-by-row to a serial register and then read out one column at a time in the floating-gate amplifier, which performs the Skipper readout step. 
The CCD has four readout amplifiers that each read out in parallel a ``quadrant'' of 512 rows by 3072 columns. 
The CCD was operated in a vacuum chamber equipped with a vacuum pump and a cryocooler, and then read out with a Low-Threshold-Acquisition board (LTA)~\cite{LTA}.

In \cref{sec:identification} we describe the measurement protocol for identifying trap candidates, and in \cref{sec:characterization} we apply selections to isolate and characterize the energy, cross-section, and lifetime of the well-behaved trap population. 
In \cref{sec:completeness} we perform a data-driven study to estimate the completeness of the measurement. 
In \cref{sec:simulation} we detail a full readout simulation using the high-energy event spectrum observed in MINOS to test the impact of the observed trap distribution on the SER. 
Finally, we conclude in \cref{sec:discussion} with a summary of the impact of the trap density on single-electron backgrounds in SENSEI.

\section{Pocket-Pumping Measurements}
\label{sec:identification}

In order to search for individual traps in this study we make use of the pocket-pumping technique described in detail in~\cite{hall_situ_2017,trapsOscura}.
This method requires an approximately flat field of charge in the active area of the CCD. While typically this flat field is generated using an external light source, such light sources are not available for in-situ calibration of CCDs used in rare-event searches.
Instead, in this study we make use of the ``pixel-pumping'' technique described in~\cite{wu_characterization_2026}.
In this technique a controlled amount of spurious charge is generated in the active area by clocking the vertical gate voltages to shift charge back and forth between adjacent pixels.
In this work we used a voltage swing of 10.25~V and 200,000 shifts to generate approximately $2000~e^-$ of charge per pixel, which serves as the flat field of charge for pocket-pumping. 
After generating this flat field, we pocket pump, which refers to asymmetric pixel pumps between the three pixel phases with varying phase times. Here a pump is defined as one complete cycle shifting charge between contiguous phases in the same pixel. 
If a trap is present in an edge phase, charge may be captured, released later, and spill into the neighboring pixel; repeating this cycle for a phase near the resonance of the trap emission timescale $\tau_{\mathrm{e}}$ produces a ``dipole'' signal relative to the uniform background. 
We refer to our edge phases as $\phi_1$ and $\phi_3$, and $\phi_2$ is the phase under which charge is held during an exposure. We use the same optimization described in~\cite{trapsOscura}, where charge is held in $\phi_2$ for $t_{\mathrm{ph}}$ and held in $\phi_3$ and $\phi_1$ for $7 t_{\mathrm{ph}}$, which maximizes the trap intensity while minimizing the pumping sequence time. 

The number of charge carriers trapped and released by pocket-pumping manifests as the intensity of the dipole ($I{(t_{\mathrm{ph}})}$), which is defined as half the difference in charge between the two pixels. 
This can be modeled using Shockley--Read--Hall theory~\cite{PhysRev.87.835,PhysRev.87.387}, where $I(t_{\mathrm{ph}})$ depends on the number of pumps $N_{\mathrm{pumps}}$, the probability of capture $P_{\mathrm{c}}$, the trap pumping ``effectiveness" $D_{\mathrm{t}}$, and the emission probability $P_{\mathrm{e}}$. This gives:

\begin{align}\label{eq:I_fit}
    I(t_{\mathrm{ph}}) =N_{\mathrm{pumps}}D_{\mathrm{t}} P_{\mathrm{c}} \left(e^{-\frac{t_{\mathrm{ph}}}{\tau_{\mathrm{e}}}}-e^{-8\frac{t_{\mathrm{ph}}}{\tau_{\mathrm{e}}}}\right)
\end{align}

The measurement procedure was as follows: for a given temperature, charge-generating shifts were performed as described above to uniformly illuminate the CCD with charge. 
Then the pocket-pumping protocol was performed with $N_{\mathrm{pumps}}=3000$ to obtain well-defined non-saturated dipoles for various $t_{\mathrm{ph}}$. 
For most temperatures, $t_{\mathrm{ph}}$ was varied from approximately $5\times 10^{-5}$~s to $0.167$~s for a total of 18 images per temperature. For several high-temperature points this was extended to $1.03$~s with an additional 7 images per temperature. 
Only one Skipper sample was taken to reduce readout time; the associated readout noise is approximately $3$--$4e^-$. Gain was calibrated using a nominal constant because the results are not sensitive to the exact value. 
Temperature was varied over 125--210~K.
We specifically took longer $t_{\mathrm{ph}}$ measurements at higher temperatures to maximize sensitivity to longer trap emission times. 
The exact pumping sequence is the same as described in~\cite{Brusco:2025kus}.

Dipoles are identified using the following algorithm.
First, we subtract the median charge from each row $r$. Then, we multiply the corrected pixel charge in row $r$ ($x_r$) by the corrected pixel charge in the same column of the previous row ($x_{r-1}$).
If a dipole is present, this computation yields a negative value. For a candidate $x_r$ we require:

\begin{align*}
    x_r x_{r-1} < -(3 \sigma_{\mathrm{MAD}})^2
\end{align*}

where $\sigma_{\mathrm{MAD}}$ is the noise level of the pixel with respect to the median absolute deviation (MAD) of pixels in this row. The median is used here to get a measure of the noise that is insensitive to outliers. This results in 9333 candidate dipoles.

We then log the dipole intensity of each candidate across all probed $t_{\mathrm{ph}}$ values at a single temperature, discarding any dipole that appears in only one image. This then forms a candidate catalog containing intensity $I(t_{\mathrm{ph}})=\frac{x_r - x_{r-1}}{2}$ for every temperature.

The uncertainty on each dipole-intensity measurement was determined empirically using 3000 control pixel pairs per quadrant. We chose coordinates outside a 20-pixel radius surrounding detected candidates and calculated the intensity after row-median subtraction separately for each temperature, quadrant and phase delay. We defined the uncertainty as half the 16th--84th percentile width of the centered control pair intensity distribution.

\section{Trap Characterization}
\label{sec:characterization}

The trap candidates for a given temperature are then fit with \cref{eq:I_fit} plus a pedestal term $I_0$. In practice we do not fit $P_{\mathrm{c}}$ and $D_{\mathrm{t}}$ separately; they are absorbed into a fit coefficient term $A=N_{\mathrm{pumps}}D_{\mathrm{t}} P_{\mathrm{c}}$. Thus \cref{eq:I_fit} becomes:
\begin{align}\label{eq:i_functional_form}
    I(t_{\mathrm{ph}}) = A \left(e^{-\frac{t_{\mathrm{ph}}}{\tau_{\mathrm{e}}}}-e^{-8\frac{t_{\mathrm{ph}}} {\tau_{\mathrm{e}}}}\right) + I_0
\end{align}

The pedestal term accounts for a $t_{\mathrm{ph}}$-independent dipole component that we observe at the trap coordinates in data, which grows from a negligible level at 150~K to $\mathcal{O}(700)\,e^-$ above 175~K, and can result in dipoles appearing asymmetric, with one significantly bright pixel next to a dim but not dark pixel, as opposed to a typical dipole with approximately equal charge in each lobe. We hypothesize that this is due to neighboring pixels generating dark current at different rates at high temperatures. 

The initial finder criterion was intentionally loose to identify asymmetric dipoles, but as a result, will often detect neighboring pixels of these lobes as additional dipoles. After fitting the intensity curves we identify these duplicates and retain a single representative coordinate for each component. This reduces the candidate count to 6809 de-aliased trap candidates. Successfully fitted trap candidates are then filtered with several selections to identify traps that behave according to the Shockley--Read--Hall theory.

First, the trap must have a good fit at a given temperature. We require the fitted amplitude $A$ to be at least 3 times its fit uncertainty: $\frac{|A|}{\sigma_A} \ge 3$ to reject signals with weak pumping strength. The fitted emission time $\tau_{\mathrm{e}}$ must also have a relative fit uncertainty $\frac{\sigma_{\tau_{\rm e}}}{\tau_{\rm e}}< 0.5$ to reject curves whose peak is insufficiently constrained by the measured delay range. Goodness-of-fit is then evaluated using a reduced $\chi^2$ test statistic. We require $\chi^2_{\mathrm{red}} < 3$. We also require $\Delta \chi^2$, defined as $\chi^2_{\mathrm{flat\,fit}} - \chi^2_{\mathrm{intensity\,fit}}$ to exceed a temperature-dependent threshold chosen such that 0.1\% of pure noise locations would pass this selection.  This threshold is determined with a calibrated detection table built by sampling thousands of trap-free pixel pairs ($\sim2000$ per quadrant), running the identical trap intensity fits and then computing $\Delta \chi^2$ for every temperature. 

In order to characterize the energy and cross-section of the identified populations, traps must appear at multiple temperatures. If a trap does not have good intensity fits for at least four different temperatures, it is discarded. We also notice traps ($\sim400$) that switch orientation and have opposite-sign intensity fits for some but not all temperatures, which are indicative of dual-response traps; these traps are also discarded. These selections reduce the catalog to 2587 trap candidates, and give us the trap emission time $\tau_{\mathrm{e}}$ as a function of temperature.

The emission time $\tau_{\mathrm{e}}$ for a trap can be modeled as a function of lattice temperature $T$, the thermal velocity of the charge carriers $v_{\mathrm{th}}$, the effective density of states in the valence band $N_{\mathrm{v}}$, the trap depth above the valence-band edge $E_{\mathrm{t}}-E_{\mathrm{v}}$ [eV], and the trap cross-section $\sigma$ [cm$^2$]~\cite{PhysRev.87.835,trapsOscura,HallPc,Brusco:2025kus}. 
Defining $E\equiv E_{\mathrm{trap}}\equiv E_{\mathrm{t}}-E_{\mathrm{v}}$, this gives:

\begin{align}\label{eq:tau(T)}
    \tau_{\mathrm{e}} = \frac{1}{\sigma v_{\mathrm{th}}N_{\mathrm{v}}}e^{\frac{E_{\mathrm{trap}}}{k_{\mathrm{B}} T}},
\end{align}
where the temperature-dependent $v_{\mathrm{th}}$ and $N_{\mathrm{v}}$ are defined as

\begin{align}
    v_{\mathrm{th}} = \sqrt{\frac{3k_{\mathrm{B}} T}{m_{\mathrm{cond}}}}~~~\text{and}~~~
    N_{\mathrm{v}} = 2 \left[ 2\pi m_{\mathrm{dens}} \frac{k_{\mathrm{B}} T}{h^2}\right]^{3/2}.
\end{align}

\noindent Here, $h$ is Planck's constant, $m_{\mathrm{cond}} \simeq 0.41\,m_{\mathrm{e}}$ and $m_{\mathrm{dens}} \simeq 0.94\,m_{\mathrm{e}}$ are the hole effective masses for conductivity and density of states, respectively, between 100~K and 200~K~\cite{GreenCon}, and $m_{\mathrm{e}}$ is the electron rest mass.

\begin{figure}[!htb]
    \centering
    \includegraphics[width=\linewidth]{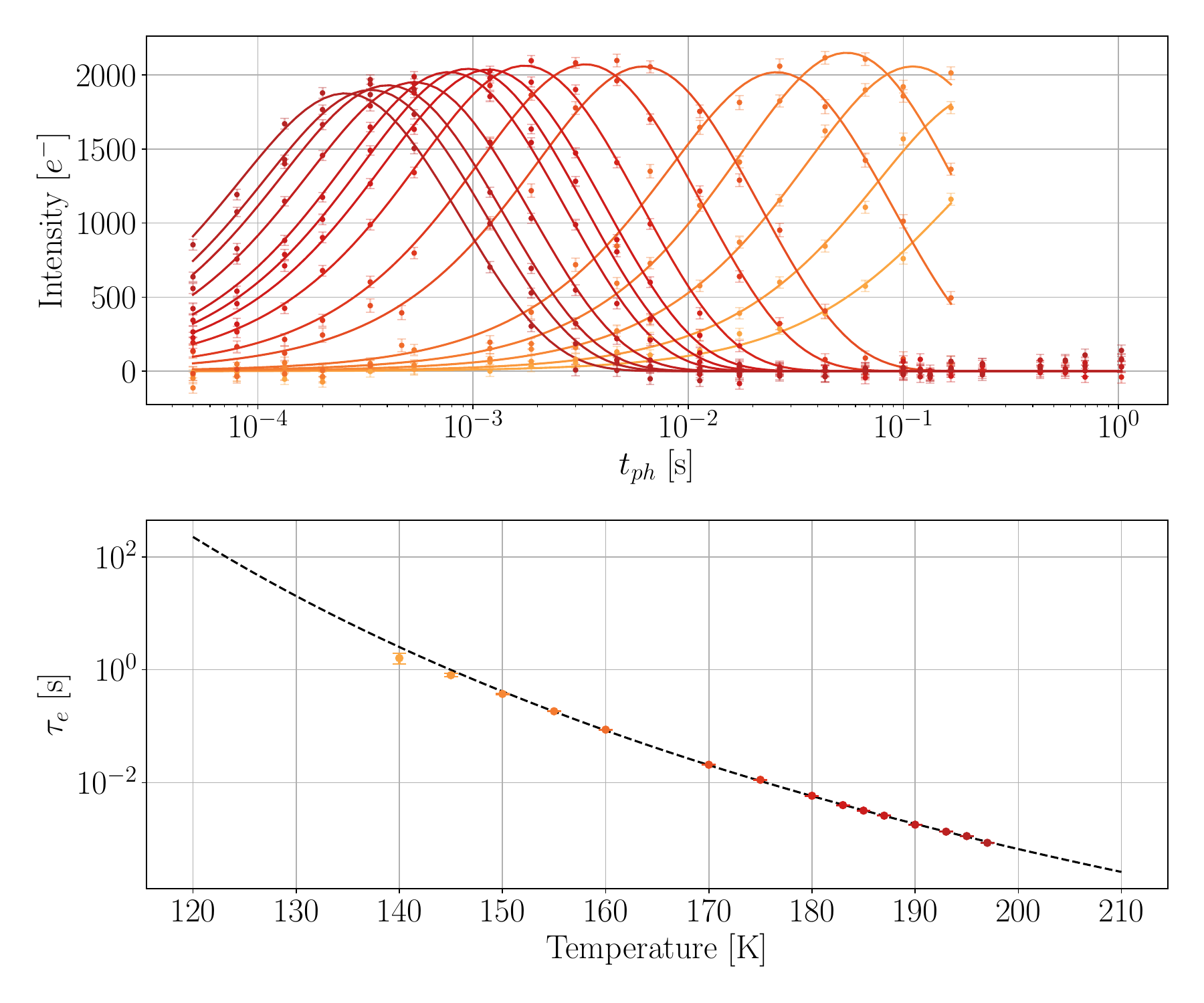}
    \caption{Example of a characterized charge trap in quadrant 1 of our CCD. The \textbf{top} panel shows the intensity (in units of electron charge) as a function of $t_{\mathrm{ph}}$ fitted according to~\cref{eq:I_fit} after removing the pedestal term. The \textbf{bottom} panel shows $\tau_{\mathrm{e}}$ as a function of temperature for the same trap with the fitted line from~\cref{eq:tau(T)}. The points are colored by temperature. }
    \label{fig:good_trap}
\end{figure}

For each trap candidate, we track the fitted $\tau_{\mathrm{e}}$ across all temperatures. We then fit these measurements with \cref{eq:tau(T)}, and for the traps that are fitted successfully we apply a $\chi^2$ fit test. Here we require $\chi^2_{\mathrm{red}} < 10$. To ensure the traps we characterize are accurately measured, we require the one-standard-deviation interval $[\sigma_-, \sigma_+]$ on the fitted cross-section to lie within a factor of two of the best-fit value $\hat\sigma$: $\hat\sigma/2 < \sigma_-$ and $\sigma_+ < 2\hat\sigma$. This results in a final characterized trap count of \ntraps\;traps. An example is shown in \cref{fig:good_trap}. The full cutflow is shown in \cref{tab:trap_cutflow}. 

\begin{table}[t]
    \centering
    \caption{Trap-candidate cutflow. Counts are cumulative
    after each selection.}
    \label{tab:trap_cutflow}
    \begingroup
    \renewcommand{\arraystretch}{1.2}
    \begin{tabular}{@{}lr@{}}
        \toprule
        Selection & Candidates \\
        \midrule
        Initial dipole candidates          & 9\,333 \\
        Duplicate removal                  & 6\,809 \\
        Temperature and orientation cuts   & 2\,587 \\
        Energy/$\sigma$ fit selections      & 1\,787 \\
        \bottomrule
    \end{tabular}
    \endgroup
\end{table}

\begin{figure}[!htb]
    \centering
    \includegraphics[width=\linewidth]{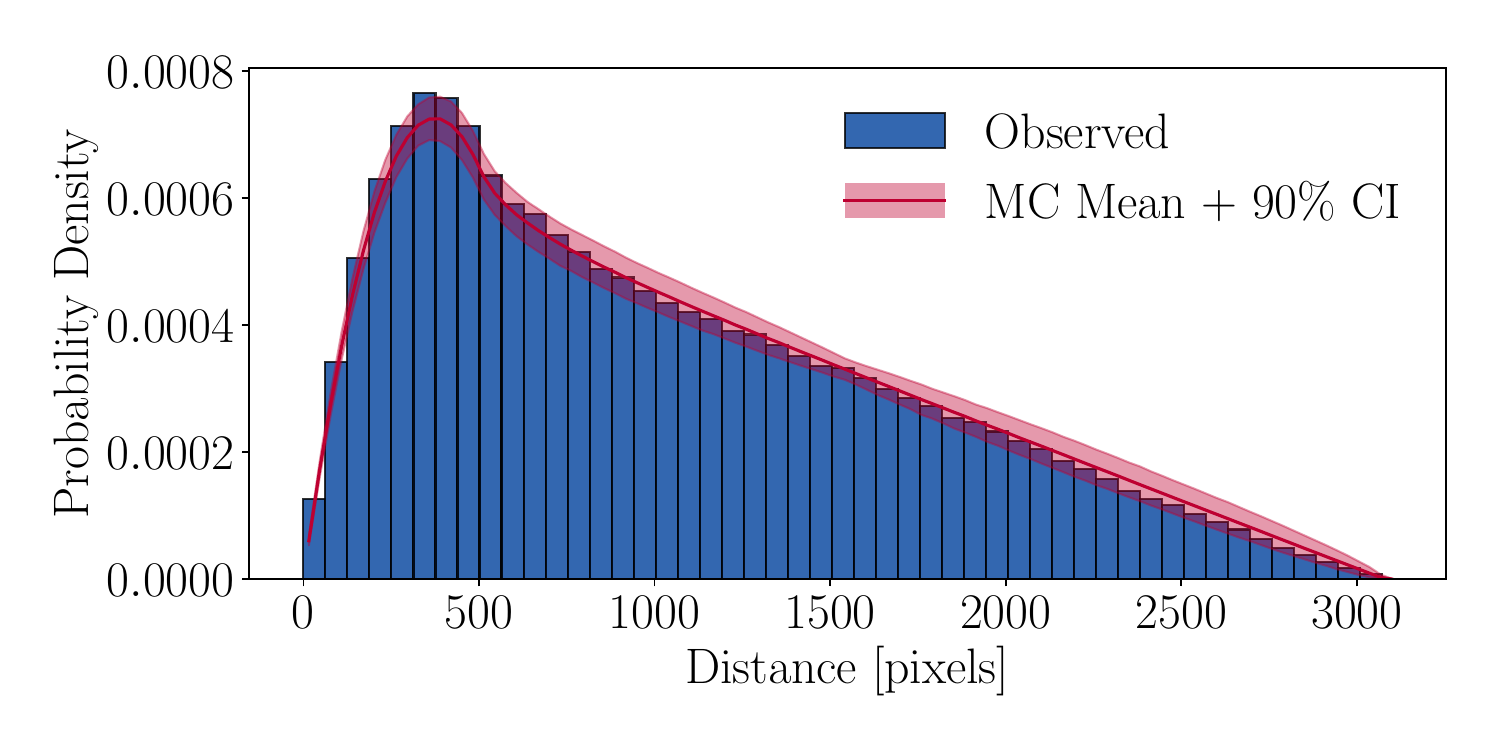}
    \caption{Probability density of the distance between all characterized traps, combining all four quadrants. The red curve and shaded band show the mean and 90\% confidence interval from $10^4$ toy Monte-Carlo simulations assuming a spatially uniform trap distribution. We see reasonable agreement between the simulation and the data.}
    \label{fig:trap_distribution}
\end{figure}

We analyze the spatial distribution of these characterized traps using toy Monte-Carlo simulations. 
$10^4$ simulations were performed under the assumption of spatial uniformity. \cref{fig:trap_distribution} shows the histogram of the distances between all traps for all quadrants combined, along with the simulation result and its corresponding 90\% confidence band assuming the traps are distributed uniformly. 
The characterized trap population appears consistent with an approximately uniform distribution.

\begin{figure}[!htb]
    \centering
    \includegraphics[width=\linewidth]{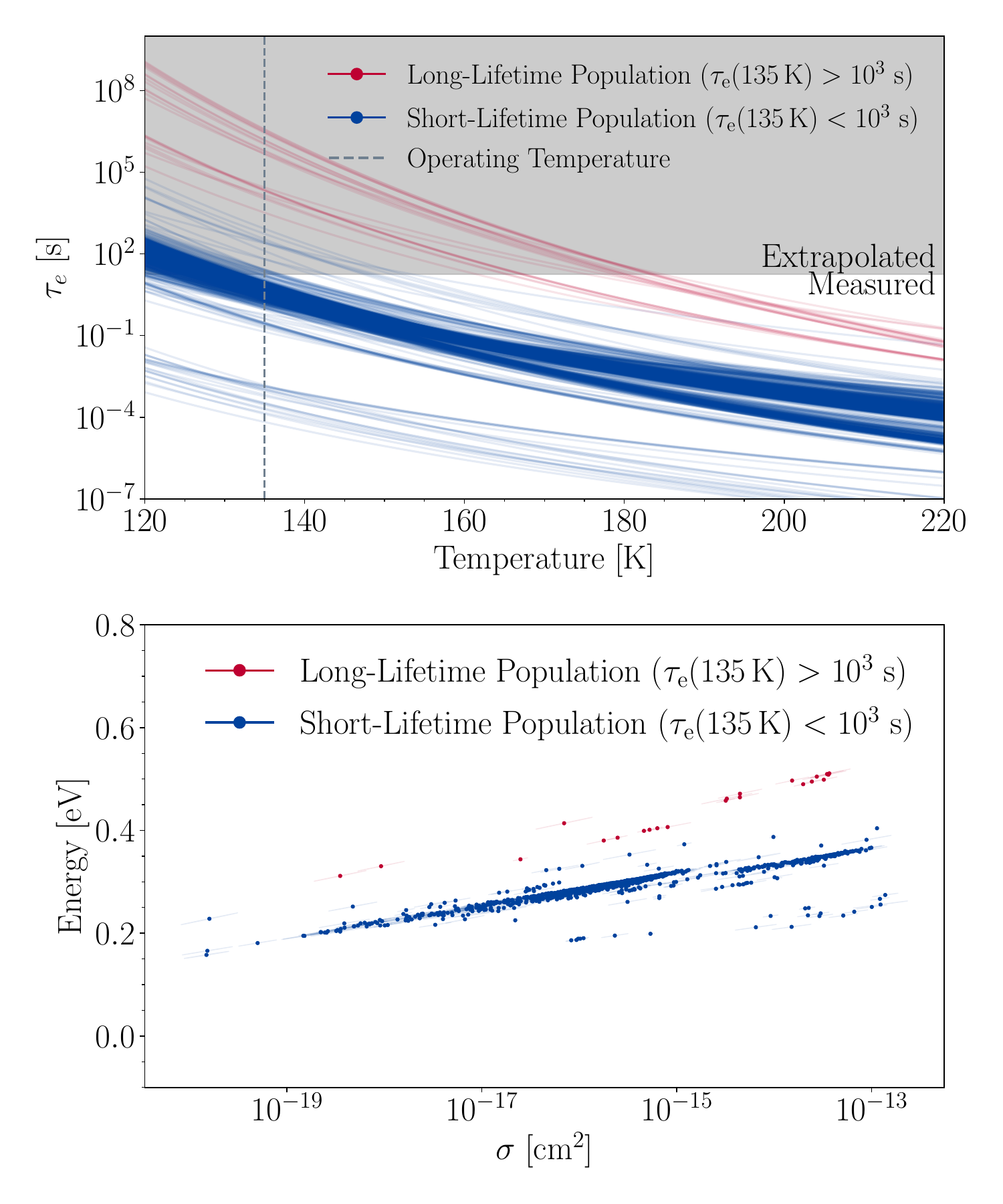}
    \caption{All \ntraps\;characterized traps, colored by observational group: ``Long-Lifetime" (red; mostly $>0.4$~eV) and ``Short-Lifetime" (blue; mostly $\sim0.2$--$0.3$~eV). \textbf{Top}: Emission time $\tau_{\mathrm{e}}$ as a function of temperature. The shaded gray region marks the extrapolated regime where no trap has a direct measurement at any temperature. The dashed vertical line marks the SENSEI@SNOLAB operating temperature of 135~K. \textbf{Bottom}: Fitted trap energy $E_{\mathrm{trap}}$ versus capture cross-section $\sigma$ for the same traps.}
    \label{fig:all_traps}
\end{figure}

The measured energy and cross-section for our characterized trap population is presented in \cref{fig:all_traps}. 
The shaded gray region begins where we no longer have direct measurements of $\tau_{\mathrm{e}}$ for any trap; all lines in this region are extrapolated from fitting with \cref{eq:tau(T)}. 
The white region denotes the measured trap emission times. The underground environment at MINOS shields the CCD from cosmic events that destroy the flat charge field and allows us to take data with longer delay times, thereby allowing us to probe significantly higher energies and cross-sections than were accessible previously~\cite{Brusco:2025kus}.
We observe a large population centered around 0.2~eV, dubbed the ``Short-Lifetime'' population and a smaller secondary population centered around 0.4--0.5~eV, which we dub the ``Long-Lifetime'' population. The latter population was not accessible in previous measurements.

\begin{figure}
    \centering
    \includegraphics[width=\linewidth]{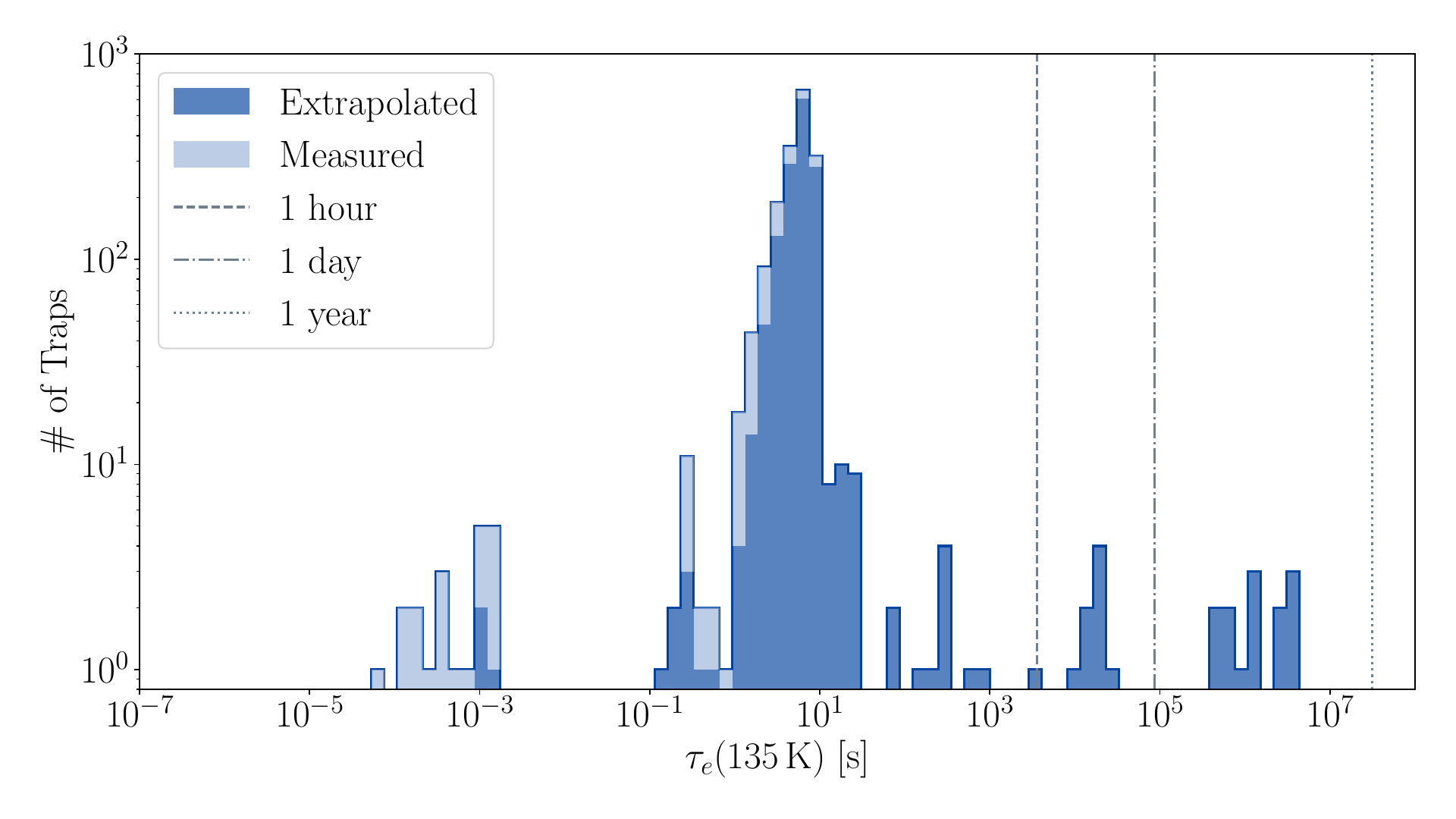}
    \caption{ Histogram of the predicted emission time $\tau_{\mathrm{e}}$ at 135~K for all \ntraps\;well-characterized traps. Light blue bars indicate the portion of the total population for which $\tau_{\mathrm{e}}$ was directly measured; dark blue bars indicate traps extrapolated using \cref{eq:tau(T)}.}
     \label{fig:tau_at_135k}
\end{figure}

We then use \cref{eq:tau(T)} to extrapolate the characteristic emission times for all characterized traps at 135~K, the typical operating temperature of SENSEI@SNOLAB~\cite{SENSEI:2023zdf}.  This distribution is shown in \cref{fig:tau_at_135k}. Not all traps have direct $\tau_{\mathrm{e}}$ measurements at every temperature. At temperature $T$, a trap's emission time is classified as \textit{measured} if the intensity was successfully fit at that temperature and as \textit{extrapolated} if the emission time is obtained solely from the fit to \cref{eq:tau(T)}.

\section{Measurement Completeness}
\label{sec:completeness}

To understand hypothetical trap populations probed by this measurement we study our measurement protocol with simulation. To do this, we define a probability of characterization $P_{\mathrm{char}}(E,\sigma)$ which is given by the probability that a trap of given energy and cross-section enters our finder, passes our intensity fit requirements for at least four temperatures, and passes the energy fit and cross-section precision requirements. This probability was calculated with a Monte-Carlo injection procedure.

A true trap has a specified energy $E$ and cross-section $\sigma$. Using \cref{eq:tau(T)} the emission time is specified for every temperature. For each phase delay $t_{\mathrm{ph}}$ the emission rate is simply $\lambda_{\rm e} = 1/\tau_{\mathrm{e}}$, and the probability of emission is given based on our measurement protocol and is defined in \cref{eq:I_fit}. In the same equation $P_{\mathrm{c}} D_{\mathrm{t}}$ is absorbed into a single amplitude when fitting the measured intensity curves. However, for a true trap $P_{\mathrm{c}}$ is set by the capture and occupancy evolution during the pumping sequence, which is based on the capture rate $\lambda_{\mathrm{c}}$. 

\begin{align}
    \label{eq:capture_rate}
    \lambda_{\mathrm{c}} = \frac{\sigma v_{\mathrm{th}}(T) n_{\mathrm{e}} }{V_{\mathrm{p}}}
\end{align}

where $n_{\mathrm{e}}$ is the number of electrons in a charge packet and $V_{\mathrm{p}}$ is the effective volume of the charge packet. This latter parameter is estimated using the product of a single carrier's lateral and vertical extent inside the potential well of the pixel. The in-plane packet area is approximated as the product of the $12~\mu\mathrm{m}$ active-channel width and the $5~\mu\mathrm{m}$ length of one pixel phase~\cite{Holland:2003,wu_characterization_2026}. The vertical extent is estimated by the thermal (Boltzmann) spread of a carrier in the buried-channel potential well $\sigma_z=\sqrt{\varepsilon_{\mathrm{Si}}k_BT/(q^2N_{\mathrm{A}})}\approx24$~nm using a channel doping $N_{\mathrm{A}}\approx1.3\times10^{16}~\mathrm{cm}^{-3}$ obtained from the implant dose~\cite{Holland:2003} and buried-channel depth~\cite{wu_characterization_2026}. 
This gives $V_{\mathrm{p}}\approx1~\mu\mathrm{m}^3$, which we use as the representative value for our results here. 

The steady-state occupation probability $p_\infty$ while the charge packet is present is given by 
\begin{align}
    p_\infty = \frac{\lambda_{\mathrm{c}}}{\lambda_{\mathrm{c}}+\lambda_{\mathrm{e}}}
\end{align}
and thus the exact mean occupancy over $N_{\mathrm{pumps}}$ is 
\begin{align}
    \label{eq:P_eff}
    P_{\mathrm{eff}}(t_{\mathrm{ph}}) ={}& p_\infty \left(1-e^{-7 t_{\mathrm{ph}}(\lambda_{\mathrm{c}}+ \lambda_{\mathrm{e}})}\right) \nonumber \\
    &\times \frac{1}{N_{\mathrm{pumps}}} \sum_{k=0}^{N_{\mathrm{pumps}} -1}\left(N_{\mathrm{pumps}} - k\right) \nonumber \\
    &\times \left(e^{-7 t_{\mathrm{ph}}(\lambda_{\mathrm{c}}+ \lambda_{\mathrm{e}})} e^{-9 t_{\mathrm{ph}}\lambda_{\mathrm{e}}}\right)^k
\end{align}

This expression accounts for incomplete capture, emission between cycles, and the buildup of occupancy across the full pumping sequence. It approaches 1 when capture saturates and is approximately proportional to $\sigma$ in the weak-capture limit. 

For a fixed $(E,\sigma,T)$ the predicted physical dipole waveform is:
\begin{align}
    I_{\mathrm{phys}}(t_{\mathrm{ph}}) ={}& o D_{\mathrm{t}} N_{\mathrm{pumps}} P_{\mathrm{eff}} \nonumber \\
    &\times \left[e^{-t_{\mathrm{ph}}/\tau_{\mathrm{e}}} - e^{-8 t_{\mathrm{ph}}/\tau_{\mathrm{e}}}\right]
\end{align}
where $o=\pm1$ is the orientation of the dipole. Therefore, for fixed $(E,\sigma)$ the only remaining trap-specific parameter is the dimensionless pumping ``effectiveness'' parameter $D_{\mathrm{t}}$, which scales the dipole amplitude. This captures variation in the signal, and is independent of $(E,\sigma)$. We observe $D_{\mathrm{t}}$ of the characterized population to peak strongly at 1. We observe a tail of $D_{\mathrm{t}}$ extending downwards to $\sim0.2$, which is the lower bound of our measurement sensitivity, and therefore suspect there are likely traps with even smaller $D_{\mathrm{t}}$. Traps with small $D_{\mathrm{t}}$ have non-ideal capture and/or emission for some reason; we hypothesize that these could be traps on the edges of the phases and therefore see less effective charge or perhaps traps that exist in the overlap between $\phi_2$ and an edge phase, which would lead to cancellation. With regards to the effect of these traps on the SER, the former likely captures less efficiently under normal operating conditions whereas the latter would capture just as effectively. Therefore for understanding the effect of these traps on the SER we assume $D_{\mathrm{t}} =1 $ and compute $P_{\mathrm{char}}$ for those traps.

To compute $P_{\mathrm{char}}$ we use a grid of trap energy and cross-sections and inject intensity curves into control pixel pairs in actual images. These control measurements provide the measured background fluctuations at each temperature and phase delay. We require these regions to be located outside a 20-pixel radius from an identified trap. We assign each simulated trap a quadrant and orientation shared across all temperatures. The charge seen by the trap is also estimated by using the median number of electrons $n_{\mathrm{e}}$ for a given quadrant at a given $t_{\mathrm{ph}}$ and temperature. We track each simulated trap across all temperatures, repeat the procedure for each quadrant and orientation, and then apply the same selections described in \cref{sec:characterization}. A trap is only considered ``characterized" if it successfully passes the same analysis selections applied to data.
This is repeated 100 times for a given $(E,\sigma)$ point and then averaged over quadrant and dipole orientation.

\begin{figure}
    \centering
    \includegraphics[width=\linewidth]{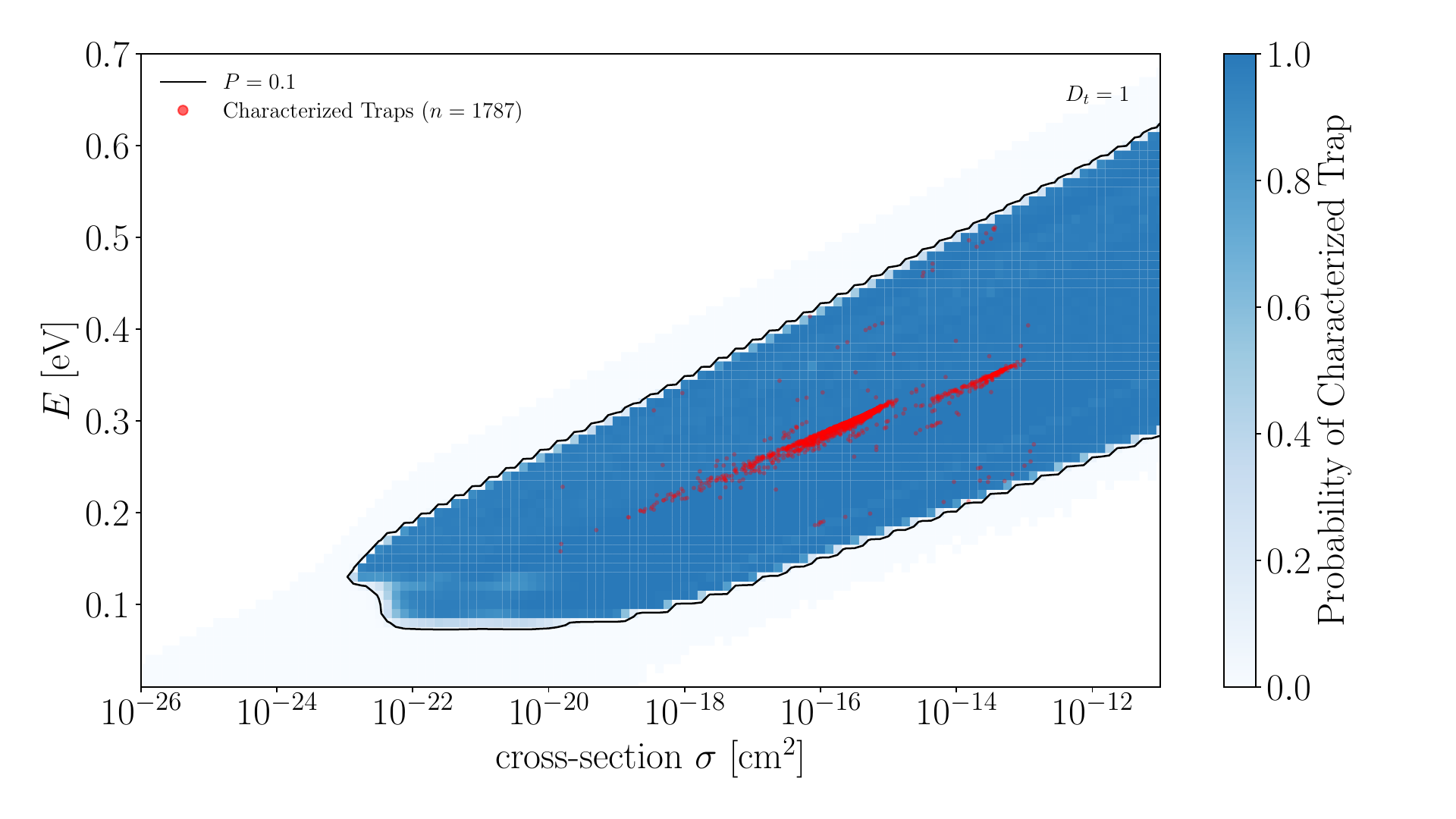}
    \caption{The probability  of characterization based on the injection curve procedure vs trap energy and cross-section is shown in blue. In red we plot the characterized trap population. Regions in white have less than a 10\% probability of being measured by the protocol used in this work. This assumes a trap effectiveness $D_{\mathrm{t}}$ of 1.}
    \label{fig:prob_of_characterization_map}
\end{figure}

This results in a 2D probability map predicting the probability of characterization in the $(E, \sigma)$ plane, and is shown in \cref{fig:prob_of_characterization_map}. As expected, the characterized traps lie in the region of high characterization probability. Of particular interest is the black border showing $P_{\mathrm{char}}$ of at least 10\%. We take this contour to be our effective boundary of measurement. This contour is dependent on the choice of $D_{\mathrm{t}}$, but is relatively robust down to $D_{\mathrm{t}}=0.5$. The effect of traps with much lower $D_{\mathrm{t}}$ on the SER is not understood and requires further study. This contour also reveals that the analysis is generally blind to high-energy and small-cross-section traps; the former because the predicted lifetime is inaccessible with the current measurement scheme and the latter because it is difficult for a small-cross-section trap to capture enough electrons to produce a measurable intensity.

 \section{Simulating Trap Emission}
\label{sec:simulation}

To investigate the effect of these traps on our SER, an end-to-end simulation architecture capable of simulating SENSEI conditions was developed. 
In the simulation, a ``CCD'' is initialized with a given charge trap density chosen to be the same as the characterized trap population (\ntraps\;traps / 6144 $\times$ 1024 pixels) and exposure-dependent and exposure-independent single-electron rates taken from the MINOS upper right (UR) quadrant in previous SENSEI results~\cite{SENSEI:2024yyt}. 
The exposure-dependent rate is taken as $4.36\times10^{-5}~e^-$/pix/day and the exposure-independent rate is taken as $9.94\times10^{-5}~e^-$/pix/image. 
For simplicity, only one quadrant is simulated (3072 columns $\times$ 512 rows). 
The traps have cross-sections and energies drawn from the distribution of the characterized traps shown in \cref{fig:all_traps}. The emission time is calculated from \cref{eq:tau(T)} assuming an operating temperature of 135~K. 
We assume an equal distribution of traps under $\phi_1$ and $\phi_3$, all initialized as empty. Schematically, a packet entering a pixel crosses $\phi_1$, and a packet leaving a pixel crosses $\phi_3$. Emission during a stationary dwell enters the trap's own pixel (held in $\phi_2$). This means for $\phi_3$ traps the emitted charge crosses the same trap on a subsequent shift and therefore can be recaptured. For traps under $\phi_1$ the charge is already downstream of the trap and therefore cannot be recaptured by the same trap. Traps under the $\phi_2$ phase were not measured in this work and therefore not simulated.

The simulation takes ``images'' by specifying an exposure time and injecting a number of single-electron counts randomly throughout the image based on Poisson fluctuations around the exposure-dependent and exposure-independent single-electron rates. 
High-energy events must also be simulated in order to match science conditions and provide the mechanism for filling traps. 
To simulate data-taking conditions, we sample a fraction of high-energy events (defined by choosing contiguous regions of positive charge with total charge $>100e^-$) from a 20~h SENSEI image taken at MINOS. 
We ignore long horizontal streaks as these correspond to high-energy events that strike the serial register during readout. 
For a chosen cluster we transplant all charge within a 60-pixel radius of pixels with charge $>100e^-$ into random non-overlapping positions in the image. 
The number of high-energy events inserted in the image scales linearly with exposure; for example, for a 4~h image, a random fifth of the total high-energy events are chosen. 
To approximate SNOLAB conditions we scale down the number of high-energy events by a factor of 10 based on the high-energy background rate measured in~\cite{SENSEI:2024yyt}. 

We then simulate readout row-by-row with delay times corresponding to two different data-taking procedures.
The first simulates $32 \times 1$ binning as in SNOLAB Run 2~\cite{SENSEI:2024yyt} ($\sim13$~min readout), and the second assumes no binning as in SNOLAB Run 1~\cite{SENSEI:2023zdf} ($\sim7$~h readout).  
Since we only model traps under edge phases $\phi_1$ and $\phi_3$, traps can only capture charge during the phase overlap window $dt\sim80~\mu$s of a row shift. 
During each gate-crossing interval we evolve the joint packet–trap state using the exact two-state transition with capture rate $\lambda_{\mathrm{c}}$ and emission rate $\lambda_{\mathrm{e}}$. This accounts for capture and emission occurring within the same transfer interval. To be conservative, we ignore any geometric parameter related to $D_{\mathrm{t}}$ (which is a pumping effectiveness term) and assume all traps capture perfectly according to the model. After an image is read out, we simulate the CCD clear used in operating conditions. We evaluate the emission probability after each row transfer dwell during readout/clearing and after every exposure.

A single trial refers to a single initialized CCD state in which 500 paired images (with and without traps, taken simultaneously) are taken and read out with ordered exposure times of 0~h, 4~h, 6~h, 10~h, and 20~h. Trap occupancy is retained between images. After every image is read out, we simulate a clear as was used in prior science runs~\cite{SENSEI:2023zdf,SENSEI:2024yyt}. The clear first clocks all charge out of the image and then applies three hours of continuous fast vertical shifts; trap occupancy is evolved throughout, and therefore charge emitted during a clear can be lost. After the images are taken we apply a set of masks developed in previous analyses~\cite{Sensei2020,SENSEI:2023zdf,SENSEI:2024yyt}.  We specifically apply the halo mask (60-pixel radius around a pixel with $>100e^-$), vertical bleed mask (all pixels above a pixel with $>100e^-$), and the hot column and hot pixel masks. For the hot column and pixel masks, which are statistical, an A/B strategy is implemented, where half the simulation set is designated as set A, and the other half as set B.  To avoid truncation bias, the set B statistics are used to decide the hot columns and pixels in A, and vice versa. This mask is applied after applying the halo and bleed masks. We perform 200 such trials for binned and unbinned readout for both MINOS and SNOLAB conditions (800 total trials).

\begin{figure}
    \centering
    \includegraphics[width=\linewidth]{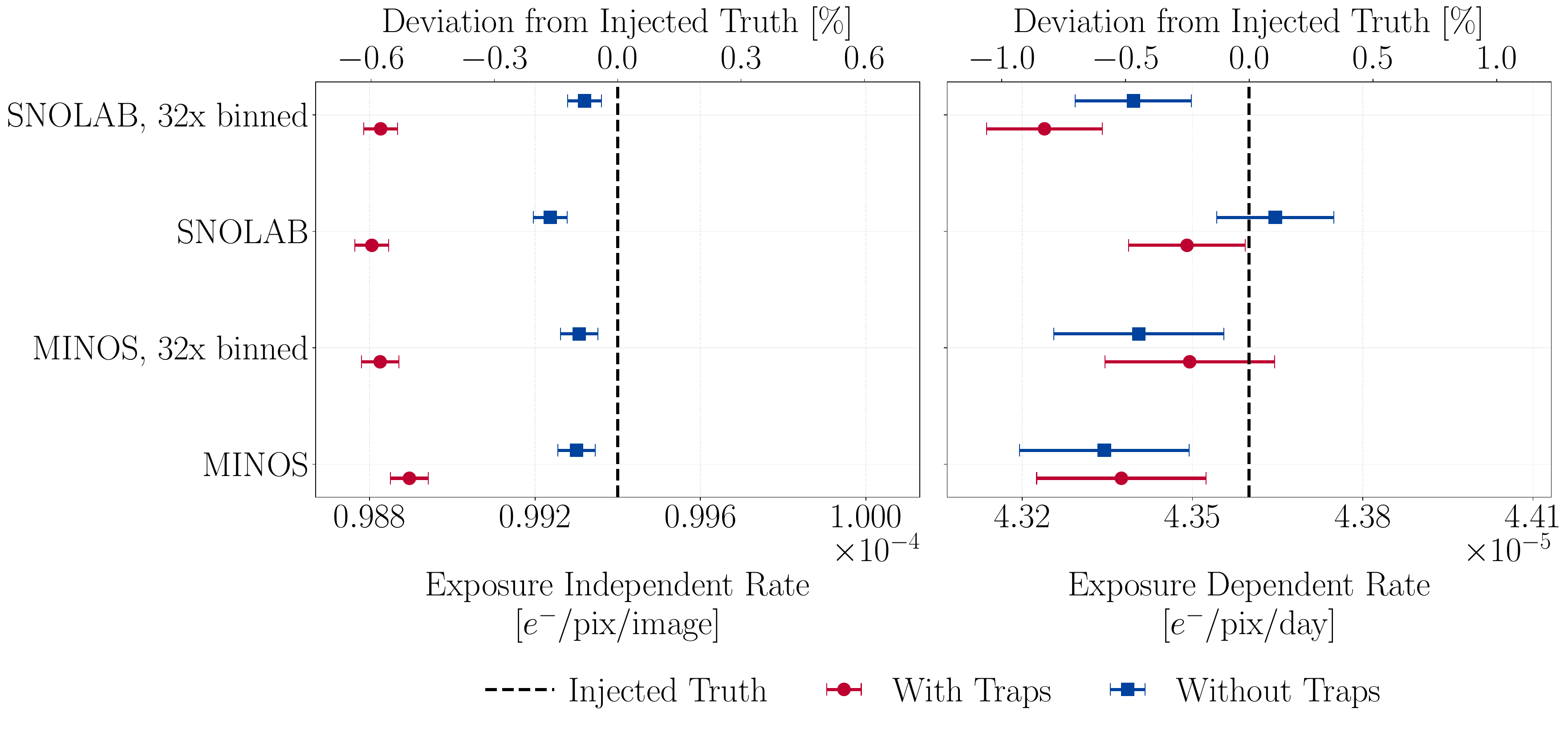}
    \caption{Exposure-independent (left) and exposure-dependent (right) rates for no-trap (blue) and trap (red) images for MINOS and SNOLAB science data-taking conditions. These assume a 3~h clear, and binned images use the same 32 row binning used in~\cite{SENSEI:2024yyt}. For the characterized population, differences are less than $1\%$, and therefore do not impact the SER. Explanations for deviations from the injected value are described in the text.}
    \label{fig:simulation_comparison}
\end{figure}

After masking, we calculate the single-electron densities for all exposures for images with and without traps. We then perform a linear fit, where the intercept gives the exposure-independent charge and the slope gives the exposure-dependent single-electron rate. The results are shown in \cref{fig:simulation_comparison}. We see the no-trap branch recovers the injected truth for the exposure-independent charge to within $0.3\%$ and the exposure-dependent rate to within $1.3\%$, providing a reasonable validation of the framework. The slight deficit observed compared to the truth value is due to coincidental overlap between single-electron injections (which are placed randomly) and other pixels already containing charge. $V_{\mathrm{p}}$ was approximated and therefore represents a systematic uncertainty. However, varying $V_{\mathrm{p}}$ by a decade in either direction in simulation only produced effects on the order of $1\%$.
We see that for the characterized trap population any deviation from the injected SER is less than 1\% compared to the no-trap images for both the exposure-dependent and exposure-independent cases, and is therefore negligible. 
We observe a $\sim0.5\%$ deficit in the exposure-independent rate compared to the no-trap scenario; this arises from traps capturing single-electron events and releasing them during the 3~h clear, where they are lost.

We then use this simulation to probe the portions of parameter space that could potentially contribute to the SER. At each point we simulate a single quadrant using a full-CCD count of 2000 traps, an amount comparable to the total characterized population, with its energies and cross-sections confined to a narrow bin around that point. 
We interpolate the simulated responses between probe centers. We assume a temperature of 135~K, $V_{\mathrm{p}}=1~\mu$m$^3$, the SNOLAB high-energy event rate and 32-row binning to compare with~\cite{SENSEI:2024yyt}. We use the same 3~h clear mentioned above. For this study, we shuffle the exposures to ensure there is no bias related to the ordering of data-taking. To ensure all contributions are only from traps, we inject zero exposure-dependent single-electron events. Then, we apply the same masks detailed earlier. 

\begin{figure}
    \centering
    \includegraphics[width=\linewidth]{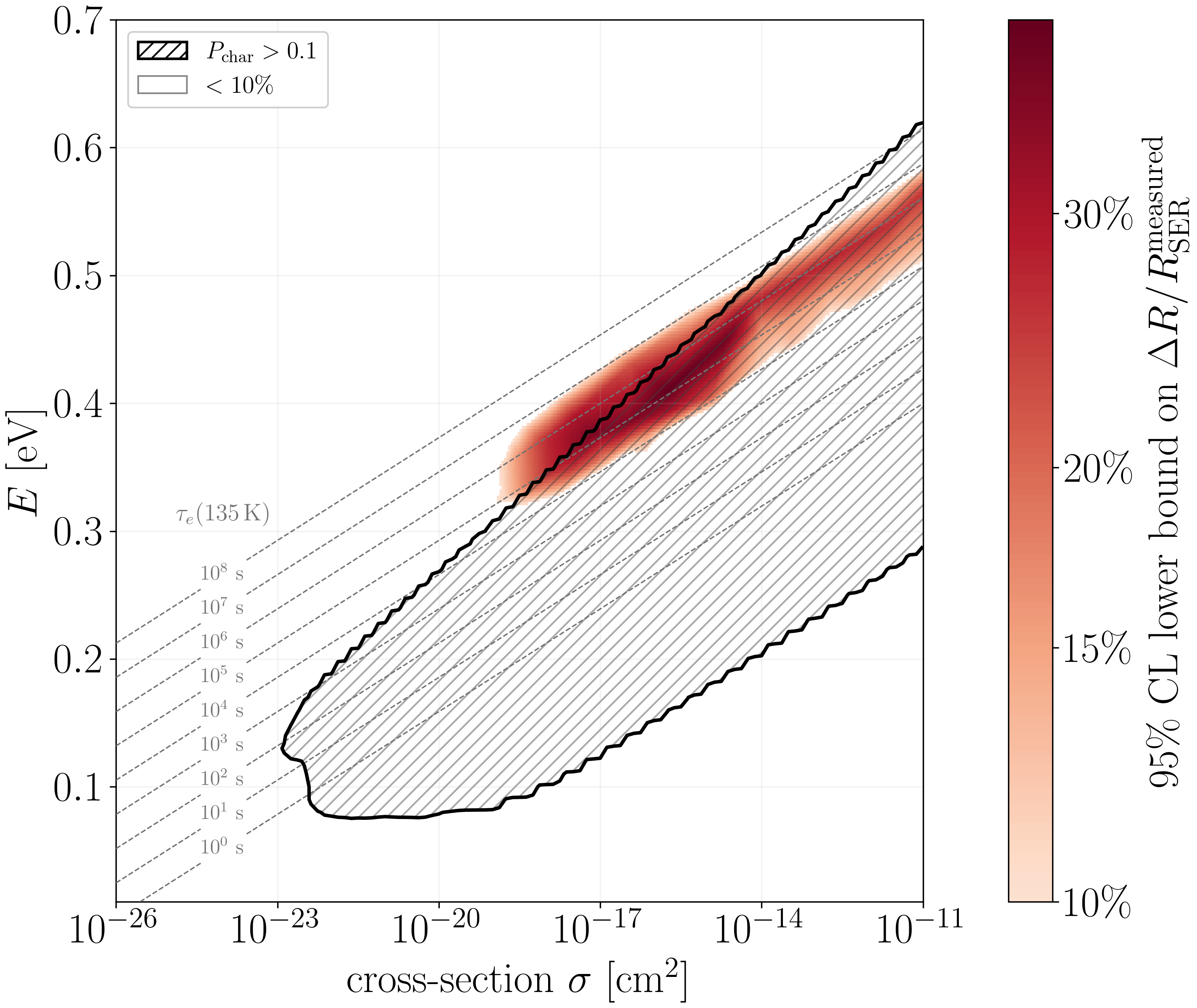}
    \caption{2000 traps concentrated within narrow regions of $(E,\sigma)$ space are simulated to study the effect on the SER. In red we show the regions of parameter space where such points contribute an excess of at least 10\% to the measured SER from SENSEI@SNOLAB. Regions in white either do not contribute or contribute $<10\%$. Shown in hatch marks is the region of parameter space corresponding to $P_{\rm char} \ge 10\%$ based on the study in \cref{sec:completeness}. A large fraction of the dangerous region is probed by this measurement, but there remain regions of parameter space that could contribute significantly. }
    \label{fig:ser_dangerous_contour}
\end{figure}

A region is classified as dangerous when the 95\% confidence lower bound on the trap-induced rate exceeds 10\% of the measured SNOLAB SER (corresponding to $1.39\times10^{-6}\,e^-$/pix/day)~\cite{SENSEI:2024yyt}. Note that this percentage is approximately linear with respect to trap count. This is shown in \cref{fig:ser_dangerous_contour}. For comparison, we also plot the region where $P_{\mathrm{char}}$ as calculated in \cref{sec:completeness} exceeds 10\%. 
The lower bound of $\tau_{\mathrm{e}}(\mathrm{135~K})\gtrsim 10^4$~s is set by the traps emitting sufficiently quickly and frequently that they are caught by the masking used in typical SENSEI analyses.
These traps either emit during the readout of a high-energy event and are caught by the hot column and/or bleed masks, or emit during the exposure following the high-energy event and do so sufficiently frequently that they are caught by the hot pixel mask. This lower bound is general. Emission rates set the upper bound of $\tau_{\mathrm{e}}(\mathrm{135~K})\lesssim 10^7$~s. Above this upper bound, traps will still contribute to the SER but the rate is too low for the simulated number of traps to have a significant impact.

Based on the completeness study, we see that the trap search performed in this work overlaps a significant fraction of the dangerous parameter space for traps that would impact the exposure-dependent SER, but it cannot fully exclude the entire dangerous population. For example, a population of traps centered around 0.4~eV and $\sigma=10^{-17}$~cm$^2$ could contribute a significant fraction of the observed rate if the concentration were large. New measurement techniques must be developed to test the existence of such trap populations. 

\section{Summary and outlook}\label{sec:discussion}

In this work we search for charge traps in a science-grade SENSEI Skipper-CCD operated underground in the MINOS cavern at Fermilab. 
Using pocket-pumping measurements performed at 23 temperatures ranging from $125$ to $210$~K, we fully characterize \ntraps\,edge-phase traps in the sensor. Due to the use of the pixel-pumping technique we were able to make these measurements in situ for the first time.
These measurements allowed us to probe traps with emission times $>10^3$~s at the SENSEI@SNOLAB operating temperature of 135~K, a regime previously inaccessible.

The characterized traps cluster predominantly at $\tau_{\mathrm{e}} \sim 10^{-1}$--$10^{2}$~s at 135~K, with a tail extending to multi-day lifetimes. 
To identify which trap populations are accessible to this measurement, we calculate $P_{\mathrm{char}}$ across $(E,\sigma)$ using data-driven injection simulations.

The directly characterized population has a very small effect on the SER, changing the injected exposure-dependent and exposure-independent rates by less than 1\%. We compare $P_{\mathrm{char}}$ with a simulation of 2000 traps concentrated within localized regions in $(E,\sigma)$ space and see that our measurement probes a significant fraction of the traps that could contribute a 10\% excess to the measured SER.  

The characterized population cannot explain the exposure-dependent single-electron rate observed in SENSEI data. While unlikely, a sufficiently large trap population in the unmeasured parameter space could contribute significantly to the exposure-dependent SER while remaining inaccessible to this measurement. Therefore, to fully rule out charge traps as a single-electron background source, different measurement techniques must be explored. Furthermore, this measurement technique is blind to traps in the center phase of the pixel. This center phase can be explored by changing which phases we pump during the measurement phase. We postpone measurements of these center phase traps to future work.

\begin{acknowledgments}
We thank James Janesick for useful discussions on spurious charge and clocking luminescence.
We are grateful for the support of the Heising-Simons Foundation under Grant No.~79921. This document was prepared by the SENSEI collaboration using the resources of the Fermi National Accelerator Laboratory (Fermilab), a U.S. Department of Energy, Office of Science, Office of High Energy Physics HEP User Facility. Fermilab is now managed by FermiForward Discovery Group, LLC, under Contract No.~89243024CSC000002.
The CCD development work was supported in part by the Director, Office of Science, of the DOE under Contract No.~DE-AC02-05CH11231.
The U.S. Government retains and the publisher, by accepting the article for publication, acknowledges that the U.S. Government retains a non-exclusive, paid-up, irrevocable, world-wide license to publish or reproduce the published form of this manuscript, or allow others to do so, for U.S. Government purposes. 
RE acknowledges support from the DOE Grant DE-SC0025309 and the Simons Investigator in Physics award MPS-SIP-00010469. AD and YW acknowledge support from the Universities Research Association and HSF Grant No.~79921. YW is also supported in part by Simons Investigator in Physics award MPS-SIP-00010469. IL and SL are supported by the Canada Foundation for Innovation (CFI) and the Province of Ontario Ministry of Research and Innovation.  
\end{acknowledgments}

%\section{Appendices}
%\appendix
%\bibliographystyle{apsrev4-1-b.bst}
\bibliography{references.bib}

\end{document}